\documentclass[aps,prb,twocolumn,superscriptaddress,longbibliography,reprint]{revtex4-2}
\usepackage{amsmath}
\usepackage{braket} % Dirac notation
\usepackage{url}
\usepackage[ruled,vlined]{algorithm2e}
\usepackage{algorithmic}
\usepackage{graphicx}% Include figure files
\usepackage{dcolumn}% Align table columns on decimal point
\usepackage{bm}% Bold math
\usepackage{amssymb}% Bold math
\usepackage{rotating}% Rotate figures and tables
\usepackage[abs]{overpic}% Over pic axis
\usepackage{xcolor}% Over pic axis color
\usepackage{booktabs}
\usepackage{tikz}
\usepackage{tabularx}% Table adjustment
\usepackage{ragged2e}
\usepackage{hyperref}
\usepackage{multirow}
\usepackage{array}
\usepackage{booktabs}
\hypersetup{colorlinks = true, citebordercolor={blue}, linkcolor={blue}, citecolor={blue}, urlcolor={blue}}

\DeclareUnicodeCharacter{2212}{-}
\DeclareUnicodeCharacter{2009}{-}

\begin{document}

%\preprint{}

\title{Constrained Shadow Tomography for Molecular Simulation on Quantum Devices}
\author{Irma Avdic}
\affiliation{Department of Chemistry and The James Franck Institute, The University of Chicago, Chicago, IL 60637 USA}
\author{Yuchen Wang}\thanks{These authors contributed equally to this work.}
\affiliation{Department of Chemistry and The James Franck Institute, The University of Chicago, Chicago, IL 60637 USA}
\author{Michael Rose}\thanks{These authors contributed equally to this work.}
\affiliation{Department of Chemistry and The James Franck Institute, The University of Chicago, Chicago, IL 60637 USA}
\author{Lillian I. Payne Torres}
\affiliation{Department of Chemistry and The James Franck Institute, The University of Chicago, Chicago, IL 60637 USA}
\author{Anna O. Schouten}
\affiliation{Department of Chemistry and The James Franck Institute, The University of Chicago, Chicago, IL 60637 USA}
\author{Kevin J. Sung}
\email{kevinsung@ibm.com}
\affiliation{IBM Quantum, IBM T.J. Watson Research Center, Yorktown Heights, NY 10598 USA}
\author{David A. Mazziotti}
\email{damazz@uchicago.edu}
\affiliation{Department of Chemistry and The James Franck Institute, The University of Chicago, Chicago, IL 60637 USA}

\date{Submitted November 12, 2025}

%Abstract

\begin{abstract}
Quantum state tomography is a fundamental task in quantum information science, enabling detailed characterization of correlations, entanglement, and electronic structure in quantum systems. However, its exponential measurement and computational demands limit scalability, motivating efficient alternatives such as classical shadows, which enable accurate prediction of many observables from randomized measurements. In this work, we introduce a bi-objective semidefinite programming approach for constrained shadow tomography, designed to reconstruct the two-particle reduced density matrix (2-RDM) from noisy or incomplete shadow data. By integrating $N$-representability constraints and nuclear-norm regularization into the optimization, the method builds an $N$-representable 2-RDM that balances fidelity to the shadow measurements with energy minimization. This unified framework mitigates noise and sampling errors while enforcing physical consistency in the reconstructed states. Numerical and hardware results demonstrate that the approach significantly improves accuracy, noise resilience, and scalability, providing a robust foundation for physically consistent fermionic state reconstruction in realistic quantum simulations.
\end{abstract}

%\maketitle
\maketitle

%Introduction
\section{Introduction}
Reconstructing a system’s quantum state from measurement data, a process known as quantum state tomography~\cite{Vogel1989, Hradil1997, James2001, Paris2004}, is foundational to quantum information science, many-body physics, and quantum chemistry. Accurate knowledge of a system’s quantum state enables quantitative characterization of correlations, entanglement, and electronic structure properties, all of which are important for performing quantum simulations, benchmarking quantum hardware, and interpreting experimental outcomes. Quantum state tomography provides the most direct route to validating state preparation, certifying entanglement in quantum devices, and probing nonclassical correlations in condensed matter and chemical systems. However, conventional full-state tomography requires measurement and computational resources that scale exponentially with system size, rendering it infeasible for all but the smallest quantum systems~\cite{Head-Marsden2020, Benavides-Riveros.2022, mcardle_2020, Smart2021_2}.

To overcome this limitation, a wide range of scalable tomographic approaches have been developed. These methods aim to infer only the most relevant or physically meaningful aspects of a quantum state rather than reconstructing the full wavefunction. Such approaches include compressed sensing~\cite{Donoho2006, Cands2006}, neural-network-based tomography~\cite{Carrasquilla2019, Mukherjee2025}, direct fidelity estimation~\cite{Flammia2011, Zhang2021, Seshadri2024}, and maximum-entropy-inspired methods based on restricted Boltzmann machines~\cite{Singh2025}, among others. A particularly powerful and widely adopted framework is that of classical shadows~\cite{Aaronson2020,Huang2020}, which enables efficient estimation of a large number of observables from randomized measurements. The shadow tomography protocol constructs a compact classical representation, termed a ``shadow'', from a set of randomized qubit or fermionic measurements, from which expectation values of many observables can be predicted with polynomial or even logarithmic scaling in system size. This paradigm has been successfully applied in a variety of contexts, including entanglement estimation~\cite{Zhang2025}, Hamiltonian learning~\cite{Castaneda2025, Somma2025, King2025}, and reduced density matrix (RDM) reconstruction for quantum chemistry~\cite{Zhao2021,Low2022}.

When specialized to fermionic systems, shadow-based methods can leverage rich physical structure, such as antisymmetry~\cite{Huggins2022}, particle-number conservation~\cite{Low2022}, spin symmetries, and Gaussian unitary transformations~\cite{Zhao2021,Zhao2024,Huang2024}, to further improve sampling efficiency and reconstruction fidelity. These fermionic shadow protocols enable scalable characterization of electronic structure properties and correlation functions relevant to molecular and solid-state systems, potentially bridging the gap between quantum simulation and classical post-processing. However, despite their theoretical efficiency, existing fermionic shadow tomography approaches face significant practical challenges. In particular, the presence of sampling noise, gate infidelity, and finite measurement statistics on quantum devices can severely degrade reconstruction accuracy.

In this work, we introduce an improved and optimized \emph{constrained shadow tomography} framework, building on prior work~\cite{Avdic2024,Avdic2024_2}, specifically tailored for the simulation of realistic fermionic systems on quantum devices. Our method formulates the reconstruction of the two-particle reduced density matrix (2-RDM) as a bi-objective semidefinite optimization, extending shadow-based tomography through a constrained formulation designed to improve robustness to noise and limited measurement data. By incorporating physical $N$-representability constraints directly into the optimization and adding a nuclear-norm regularization to the energy minimization, the method reconstructs a shadow-consistent 2-RDM that satisfies the $N$-representability conditions through an optimization that balances energy minimization against fidelity to the measured shadow data. This unified semidefinite program simultaneously mitigates errors arising from finite sampling and hardware noise while enforcing physical consistency in the resulting RDMs. Consequently, the framework can be viewed as a noise-aware, state-tailored extension of variational RDM theory~\cite{mazziotti2001, Nakata.2001, Mazziotti.2002d5q, Mazziotti.2004, Zhao.2004, Cances.2006, Gidofalvi.2008, Shenvi.2010, Verstichel.2011, M2011, Baumgratz.2012, M2012, Mazziotti.2016sve, Alcoba.2018, Mazziotti.2020z0p, Li.2021, Knight.2022, Mazziotti2023, Gao.2025dgs, Schouten.2025a9e, Piris.2021}, capable of producing physically valid density matrices under realistic quantum measurement conditions.

Through comprehensive numerical analysis, we explore how measurement requirements scale with system size, orbital basis dimension, and the choice of $N$-representability constraints, identifying practical regimes where constrained tomography yields superior performance over other approaches. Finally, we validate constrained shadow tomography through computations performed on IBM's superconducting quantum device, confirming that the advantages of constraint-enforced shadow tomography persist under realistic experimental conditions involving gate noise, readout errors, and limited sampling depth. Our results highlight the importance of integrating physical priors and optimization-based constraints into quantum state reconstruction frameworks, paving the way for accurate and scalable molecular simulation on quantum hardware.

%Theory
\section{Theory}

\subsection{Shadow tomography for fermionic systems}
Characterizing the full state of a quantum system requires exponentially scaling resources, making it infeasible for practical applications. Tomography schemes based on random measurements have emerged as more computationally accessible alternatives~\cite{Haah2017,Elben2020,Elben2022}, with classical shadow tomography being among the most widely used~\cite {Huang2020,Aaronson2020}. Classical shadow tomography allows for the reconstruction of a quantum state from repeated measurements in different unitary bases, using resources that scale logarithmically with the system size in some cases~\cite{Zhao2021}. By applying random unitary operations, $U$, to the quantum state, $\ket{\psi}$, followed by a computational basis measurement, $\ket{b}$, a ``classical shadow'' estimator can be constructed,
\begin{equation}
    \hat{\rho}_i = \mathcal{M}^{-1}\!\left(U_i^\dagger |b_i\rangle\langle b_i| U_i \right),
\end{equation}
where $\mathcal{M}$ is the measurement channel. The estimator can then be used to predict the expectation values of many observables with provable guarantees,
\begin{equation}
    \langle O \rangle_{\text{shadow}} = \frac{1}{M} \sum_{i=1}^{M} \mathrm{Tr}(O \hat{\rho}_i),
\end{equation}
which converges to $\mathrm{Tr}(O \rho)$ as the number of measurements $M$ increases. For fermionic systems, the choice of measurement unitaries is key to an optimal protocol. For example, particle-number conserving transformations or fermionic Gaussian unitaries (FGUs) preserve antisymmetry and can be implemented efficiently on qubit hardware through linear optical or Majorana representations~\cite{Huggins2022,Zhao2021,Huang2024}. These structured unitaries significantly reduce circuit depth and improve measurement fidelity compared to fully random Clifford unitaries.

The statistical efficiency of classical shadow tomography is further characterized by the shadow norm, which quantifies the variance of observable estimators under random measurements~\cite{Huang2020}. For a given observable $O$, the estimation error scales as
\begin{equation}
    \mathrm{Var}\!\left[\langle O \rangle_{\text{shadow}}\right] \leq \frac{\|O\|_{\text{sh}}^2}{M},
\end{equation}
where $\|O\|_{\text{sh}}$ is the shadow norm determined by the chosen measurement ensemble. A smaller shadow norm implies that fewer measurement samples are required to achieve a target precision, emphasizing the importance of designing measurement schemes that minimize $\|O\|_{\text{sh}}$ for low-rank or local fermionic operators.

In the context of electronic structure and many-fermion simulations, the goal is often not full state reconstruction but accurate estimation of low-order observables such as the 2-RDM. Classical shadow tomography provides an efficient statistical framework for directly estimating these quantities from measurement data, enabling integration with post-processing schemes and physical consistency.

Since fermionic wavefunctions prohibitively scale exponentially with particle number, their state tomography is often simplified to an estimation of $p$-particle reduced density matrices ($p$-RDMs). If represented in the qubit basis, estimating all $p$-RDMs on $n$ fermionic modes requires $\mathcal{O}((2n)^p/\epsilon^2)$ measurements~\cite{Bravyi2002,Jiang2020}. Using FGUs, the sample complexity can be reduced to $\mathcal{O}(\binom{\eta}{p} p^{3/2}\log(n)/\epsilon^2)$~\cite{Zhao2021}. Considering particle-number symmetry, the required number of samples can further be reduced to $\mathcal{O}(\eta^p/\epsilon^2)$~\cite{Low2022} with $\eta$ particles. Here, the estimation error remains independent of the full Hilbert space dimension.

\subsection{Constrained shadow tomography of the 2-RDM}

In fermionic shadow tomography, the conservation of parity and particle number symmetry ensures compatibility with the physically relevant symmetry sectors of many-body Hamiltonians, leading to reduced sample complexity. Similarly, ensuring the estimated $p$-RDMs correspond to $N$-particle wavefunctions, i.e., that they are $N$-representable~\cite{M2007, Coleman1963, Coleman2000, Garrod.1964, Kummer1967, Erdahl1978, mazziotti2001, Mazziotti2006_2, M2012, Mazziotti2023}, allows for a reduction in measurement overhead as previously demonstrated by some of the authors~\cite{Avdic2024}. Since the electronic Hamiltonian contains at most pairwise interactions, we restrict demonstration of the present work to 2-RDMs. The 2-RDM elements of an $N$-electron system arise from the integration of the $N$-particle density matrix over all particles except two and can be expressed as
\begin{equation}
    ^2D^{ij}_{kl} = \bra{\Psi}\hat{a}^\dagger_{i}\hat{a}^\dagger_{j}\hat{a}^{}_{l}\hat{a}^{}_{k}\ket{\Psi},
\end{equation}
where $\hat{a}_{i}^{\dagger}$ creates a particle in orbital $i$ and $\hat{a}^{}_{i}$ annihilates a particle in orbital $i$. The classical shadow representation of the 2-RDM can be expressed as
\begin{equation}
S_{n}^{pq} = \bra{\Psi}\hat{U}^{\dagger}_{n}\hat{a}^\dagger_{p}\hat{a}^\dagger_{q}\hat{a}^{}_{q}\hat{a}^{}_{p}\hat{U}_n\ket{\Psi}, \label{eq:shadow}
\end{equation}
where $\hat{U}_{n} = \exp{({\sum_{uv}A^{uv}_{n}\hat{a}^{\dagger}_{u}\hat{a}^{}_{v}})}$. Here, the indices denote spin orbitals, $A_{n}$ is a one-body anti-Hermitian matrix, and $n$ is the shadow index. Each shadow $n$ corresponds to measuring all diagonal elements of the 2-RDM after applying the one-body unitary transformation, $\hat{U}_{n}$, to the wave function. The one-body unitaries are generated by random sampling according to the Haar measure~\cite{Haar1933}, effectively rotating the orbitals into a new basis. From Eq.~(\ref{eq:shadow}), the shadow measurements can be expressed in terms of the 2-RDM elements as
\begin{equation}
S^{pq}_{n} = \sum_{ijkl}{U_{n}^{pi} U_{n}^{qj} \, {^2D^{ij}_{kl}} \, U_{n}^{ql} U_{n}^{pk}},
\end{equation}
where $U_{n} = \exp{(A_{n})}$. A sufficiently large collection of such shadows defines a system of equations that uniquely determines the 2-RDM. Consequently, this approach can be used to reconstruct the 2-RDM (or any one- or two-body expectation value) of the original quantum state. Importantly, since the measurements only involve the classical components of the 2-RDM, which commute and can therefore be measured simultaneously, the overall cost of the tomography is substantially reduced.

By defining an objective functional of the 2-RDM, ($J[^2D]$), such as the nuclear norm, Frobenius norm~\cite{Smart2022}, or the energy expectation value~\cite{Candes.2009,Cai.2010,M2011}, a convex optimization problem can be formulated with respect to the shadow constraints,
\begin{equation}
\begin{aligned}
    &\min_{^2D \in {^{N}_{2}\Tilde{P}}} \hspace{0.2cm} J[^2D] \\
    \text{such that~~} S^n_{pq} &= ((U\otimes U)^2D(U\otimes U)^T)^{pq}_{pq},
\end{aligned}
\label{eq:SDP-general}
\end{equation}
for $n \in [0,m]$, where ${^{N}_{2}\Tilde{P}}$ denotes the convex set of approximately $N$-representable 2-RDMs. For a quantum many-body system with at most pairwise interactions, minimizing the energy functional $E[^2D]$ is particularly appealing due to its direct relevance to molecular simulation. The 2-RDM can further be used to estimate the energy gradients~\cite{OBrien2019,Overy2014} and multipole moments~\cite{Gidofalvi2007}. However, to enable practical implementation on current quantum devices, the formulation must be augmented to properly account for quantum measurement errors.

Motivated by the need to consider two complementary objectives: (1) to obtain an optimal solution to a general functional of the 2-RDM and (2) to minimize measurement errors (e.g., shot/gate noise, readout errors, limited sampling depth, etc.) affecting the 2-RDM, we cast Eq.~\ref{eq:SDP-general} into the following bi-objective semidefinite-programming optimization problem~\cite{Evans1973,Zeleny1974}:
\allowdisplaybreaks
\begin{align}
    \min_{\;{}^{2}D \in {}^{N}_{2}\tilde{P}} \quad &
        J[{}^{2}D] + w \,\|{}^{2}D - {}^{2}\tilde{D}\|_* \\
    \text{subject to} \quad
        \tilde{S}^{n}_{pq} &= \big( (U \otimes U)\, {}^{2}\tilde{D}\, (U \otimes U)^{T} \big)^{pq}_{pq},
        \label{eq:SDP-dual}
\end{align}
where
\begin{align}
    {}^{2}\tilde{D} = {}^{2}D + \mathbf{E}_{1} - \mathbf{E}_{2}.
\end{align}
Here, $w>0$ is a penalty weight (i.e., regularization parameter), $\|\hspace{1.2pt}\|_*$ denotes nuclear norm, and $\textbf{E}_i\succeq0$ are positive semidefinite slack ``error'' matrices. The penalty parameter $w$ controls the trade-off between energy minimization and physical deviations of the 2-RDM from the shadow measurements. A smaller $w$ allows for a lower energy value but with potentially larger shadow-constraint violations. A similar relaxation strategy was previously employed by some of the authors to infer time-dependent quantum states from measurement data using matrix completion techniques and a modified Frobenius norm within a multiobjective SDP~\cite{Foley2012}. Related approaches have since been investigated on quantum computers using purification of the 2-RDM~\cite{Smart2022} and the Quantum Approximate Optimization Algorithm~\cite{Kotil2025}.

The second objective function in Eq.~\ref{eq:SDP-dual} minimizes the nuclear norm, or the sum of the singular values of the slack matrix, over the constraint set including the shadow constraints. The nuclear norm serves as a convex and computationally tractable surrogate for matrix rank, and efficient first-order algorithms based on singular value thresholding have been developed to solve such problems at large scale~\cite{Cai.2010}. Being convex and easy to optimize, the nuclear norm offers the best convex approximation of the rank function for matrices whose largest singular value is no greater than one~\cite{Recht2010}. When the matrix variable is symmetric and positive semidefinite, the method relaxes to the trace heuristic, i.e., $\|X\|_*=\text{Tr}(X)$, frequently used in systems and control studies~\cite{Beck1998,Mesbahi1997}.

For practical implementation, considering the energy functional of the 2-RDM and the case where $w_i=w$ for all $i$, we obtain the following bi-objective semidefinite program (SDP), which includes the relaxation for rank via the nuclear norm,
\begin{align}
\min_{^2D} \hspace{0.2cm} E[^2D] &+ w\text{Tr}(\textbf{E}_1+\textbf{E}_2)  \label{eq:SDP-shadow}\\
\text{such that} \hspace{0.2cm} ^2D &\succeq 0 \nonumber\\
^2Q &\succeq 0 \nonumber\\
^2G &\succeq 0 \nonumber\\
\textbf{E}_1 &\succeq 0 \nonumber\\
\textbf{E}_2 &\succeq 0 \nonumber\\
w &> 0 \nonumber\\
{\rm Tr}(^2D) &= N(N-1) \nonumber\\
^2Q &= f_Q(^2D) \nonumber\\
^2G &= f_G(^2D) \nonumber\\
{}^{2}\tilde{D} &= {}^{2}D + \mathbf{E}_{1} - \mathbf{E}_{2} \nonumber \\
\tilde{S}^{n}_{pq} &= \big( (U \otimes U)\, {}^{2}\tilde{D}\, (U \otimes U)^{T} \big)^{pq}_{pq}
 \nonumber.
\end{align}
The two error matrices, $\textbf{E}_1$ and $\textbf{E}_2$, act as a ``positive'' and ``negative'' direction of error, respectively, increasing and decreasing the correction to the 2-RDM simultaneously. In the limit where $w\rightarrow0$, the SDP reduces to the variational 2-RDM algorithm~\cite{Mazziotti.2004,M2011}, in which only the energy functional $E[^2D]$ is minimized subject to the $N$-representability conditions. In the limit $w\rightarrow\infty$, the SDP gains the ability to treat arbitrary quantum states, including excited and non-stationary electronic states, within the same variational framework~\cite{Liebert2025,Liebert2021}.

The present algorithm enforces a necessary subset of the $N$-representability conditions, known as the 2-positivity ($\mathrm{DQG}$) conditions~\cite{Garrod.1964, Coleman1963, Mazziotti.20015ua, M2007, M2012, Schilling2015}. The 2-positivity conditions restrict the particle-particle RDM ($^{2} D$), the hole-hole RDM ($^{2} Q$), and the particle-hole RDM ($^{2} G$) to be positive semidefinite where $f_{Q}$ and $f_{G}$ are linear maps connecting $^{2} Q$ and $^{2} G$ to $^{2} D$, respectively. These semidefinite constraints correspond to keeping the particle-particle, hole-hole, and particle-hole probability distributions nonnegative.  The 2-positivity conditions incur a computational scaling of $\mathcal{O}(r^4)$ in memory and $\mathcal{O}(r^6)$ in floating-point operations~\cite{M2011}, where $r$ denotes the number of orbitals, and therefore, constrained shadow tomography maintains overall polynomial scaling. Performing full 2-RDM tomography on a quantum device would nominally require $\mathcal{O}(r^4)$ distinct measurements, whereas our shadow-based technique reduces this requirement to approximately $n_s r^2$, where $n_s$ is the number of classical shadows and typically satisfies $n_s \ll r^2$ for a desired accuracy $\epsilon$~\cite{Avdic2024,Avdic2024_2}.

From a practical standpoint, the error-mitigating formulation presented is particularly well-suited for application on quantum devices, where finite sampling noise and gate errors can corrupt 2-RDM estimates derived from shadow tomography or other measurement schemes. By incorporating error matrices directly into variational optimization, the method provides a mechanism to regularize noisy experimental data while maintaining physical consistency through the $N$-representability constraints. Beyond molecular electronic structure, this framework is broadly applicable to quantum simulation tasks involving spin models, correlated materials, and quantum state tomography, where convex relaxation and data-driven corrections offer a scalable and physically grounded route to extracting reliable observables from imperfect measurements.

%Results
\section{Results}
To implement constrained shadow tomography, we solve the semidefinite program (SDP) in Eq.~\ref{eq:SDP-shadow} by adding shadow constraints and nuclear norm relaxation to the variational 2-RDM (v2RDM) method~\cite{Mazziotti.2004, Gidofalvi2005} in the Maple Quantum Chemistry Package~\cite{maple_2025, rdmchem_2025}. We refer to the resulting algorithm as the shadow v2RDM (sv2RDM) method. The SDP is solved using the boundary-point algorithm in Ref.~\cite{M2011}. To simulate measurement uncertainty, we generate noisy 2-RDMs by adding element-wise Gaussian perturbations to the full configuration interaction (FCI) 2-RDMs, corresponding to the Bernoulli shot noise expected from finite sampling (details can be found in the Supplemental Material Sec. I A). The Fermionic Classical Shadows (FCS) protocol is implemented as described in Ref.~\cite{Low2022} (details can be found in the Supplemental Material Sec. I B). Constrained shadow tomography is applied to the dissociation of the nitrogen dimer, ground-state energies of strongly correlated hydrogen chains~\cite{Suhai1994} with up to ten equally spaced atoms, and a rectangle-square-rectangle transition of H$_4$. The hydrogen atoms are represented in the minimal Slater-type orbital (STO-3G) basis set~\cite{Hehre1969} while nitrogen is represented in the correlation-consistent polarized valence double-zeta (cc-pVDZ) basis set~\cite{Dunning1989} in a 10~electrons-in-8~orbitals [10,8] active space. All quantum circuits used for data collection are implemented using the Qiskit package~\cite{Qiskit} and ffsim library~\cite{ffsim} on IBM’s 156-qubit superconducting processor ibm\_fez (details can be found in Supplemental Material Sec. II)\footnote{All data and sample computational script are openly available at \url{https://github.com/damazz/ConstrainedShadowTomography}.}.

\begin{figure*}
    \centering
    \includegraphics[width=\linewidth]{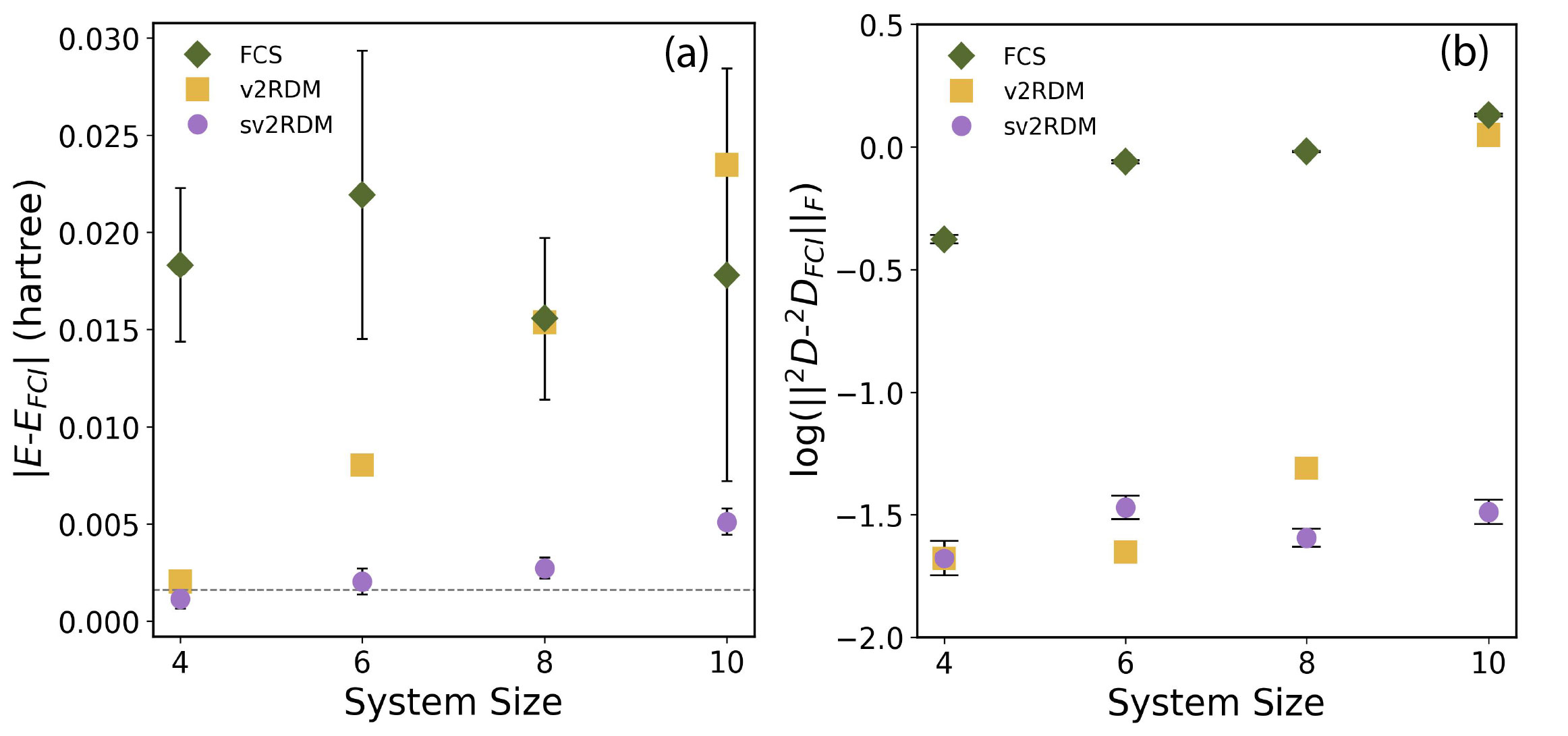}
    \caption{(a) Absolute energy error as a function of system size under a fixed total shot budget. (b) Absolute Frobenius norm of the 2-RDM error (relative to 2-RDM from FCI) plotted on a logarithmic scale. Shot budgets compared are: 16,000 unitaries × 1 shot vs. 16 unitaries × 1,000 shots for H$_4$; 36,000 unitaries × 1 shot vs. 36 unitaries × 1,000 shots for H$_6$; 160,000 unitaries × 1 shot vs. 160 unitaries × 1,000 shots for H$8$; and 300,000 unitaries × 1 shot vs. 300 unitaries × 1,000 shots for H$_{10}$. Each data point for the Fermionic Classical Shadows (FCS) and shadow variational 2-RDM (sv2RDM) methods represents the mean of 20 independent runs (10 for H$_{10}$). Error bars denote 95\% confidence intervals (approximately 2$\sigma$) computed from these measurements. All calculations are performed using the minimal basis set.}
    \label{fig:H-chains}
\end{figure*}

We begin by assessing the accuracy and consistency of the sv2RDM approach against the most-favorably scaling FCS tomography protocol~\cite{Low2022}. Figure~\ref {fig:H-chains} summarizes the dependence of absolute energy error, computed in the minimal basis set, and the Frobenius norm of the 2-RDM error (relative to the 2-RDM from full configuration interaction (FCI)) on system size under fixed total shot budgets for hydrogen chains of increasing length (H$_4$-H$_{10}$). Two sampling strategies were compared: the many-unitary/single-shot optimum for the FCS and the few-unitary/ multiple-shot protocol for the sv2RDM. Each data point represents an average over 20 independent runs (10 for H$_{10}$), with 95\% confidence intervals indicated by the error bars. Both quantum tomography methods are compared with the classical v2RDM method. Across all systems, the overall trends reveal sv2RDM outperforming FCS in both total energy and 2-RDM reconstruction with respect to the FCI results, given the same total shot budget for the two protocols. Importantly, the circuits corresponding to the shot budgets used are up to two orders of magnitude lower in depth for sv2RDM, substantially improving the feasibility of constrained shadow tomography on quantum hardware. Moreover, the sv2RDM method provides greater accuracy than the classical v2RDM method with increasing system size.

\begin{figure}[h!]
    \centering
    \includegraphics[width=\linewidth]{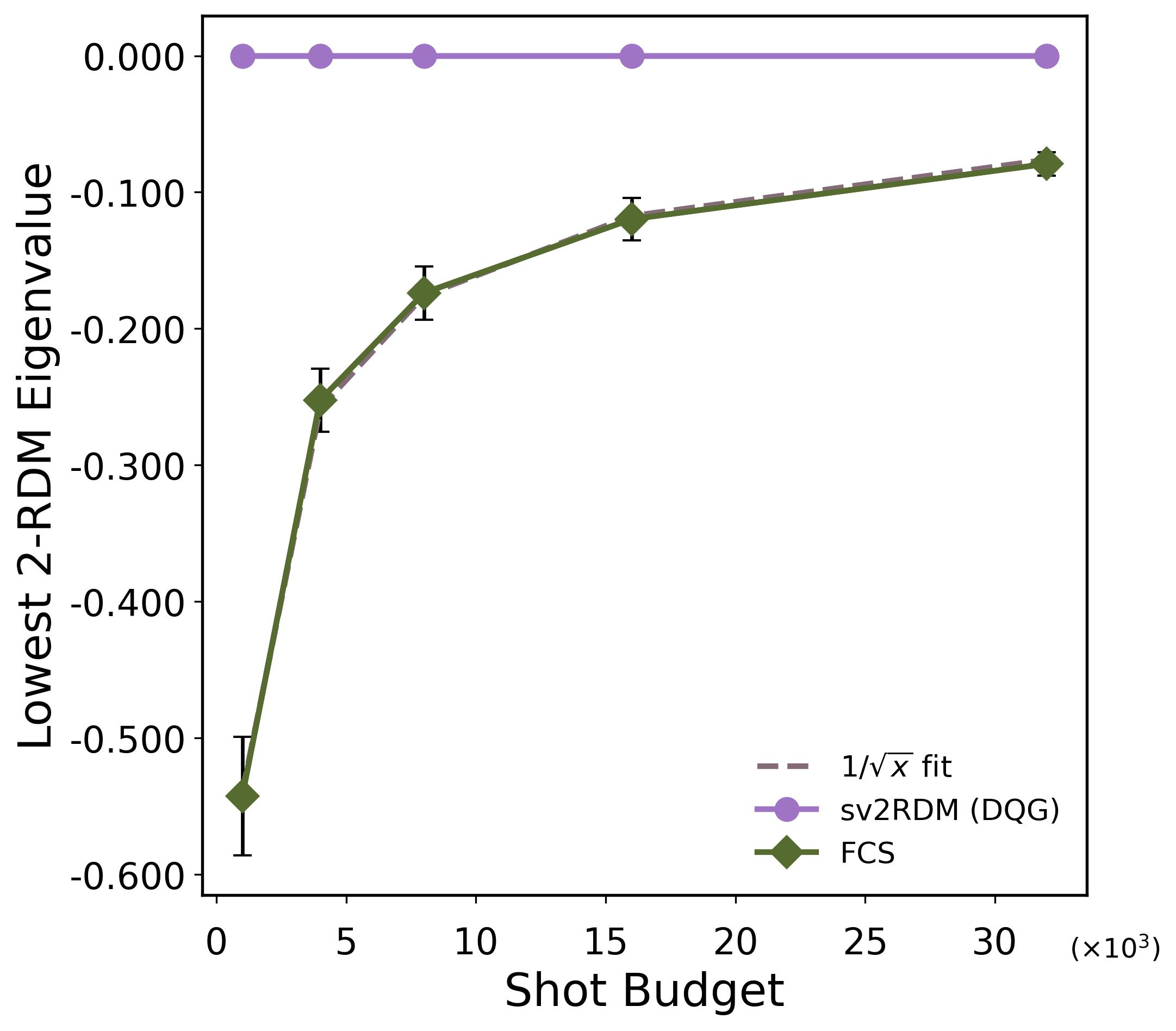}
    \caption{Lowest 2-RDM eigenvalue as a function of total shot budget for the shadow variational 2-RDM (sv2RDM) method with full 2-positivity (DQG) constraints and for the Fermionic Classical Shadows (FCS) approach applied to H$_4$ in the minimal basis set.}
    \label{fig:N-rep}
\end{figure}

\begin{figure*}
    \centering
    \includegraphics[width=\linewidth]{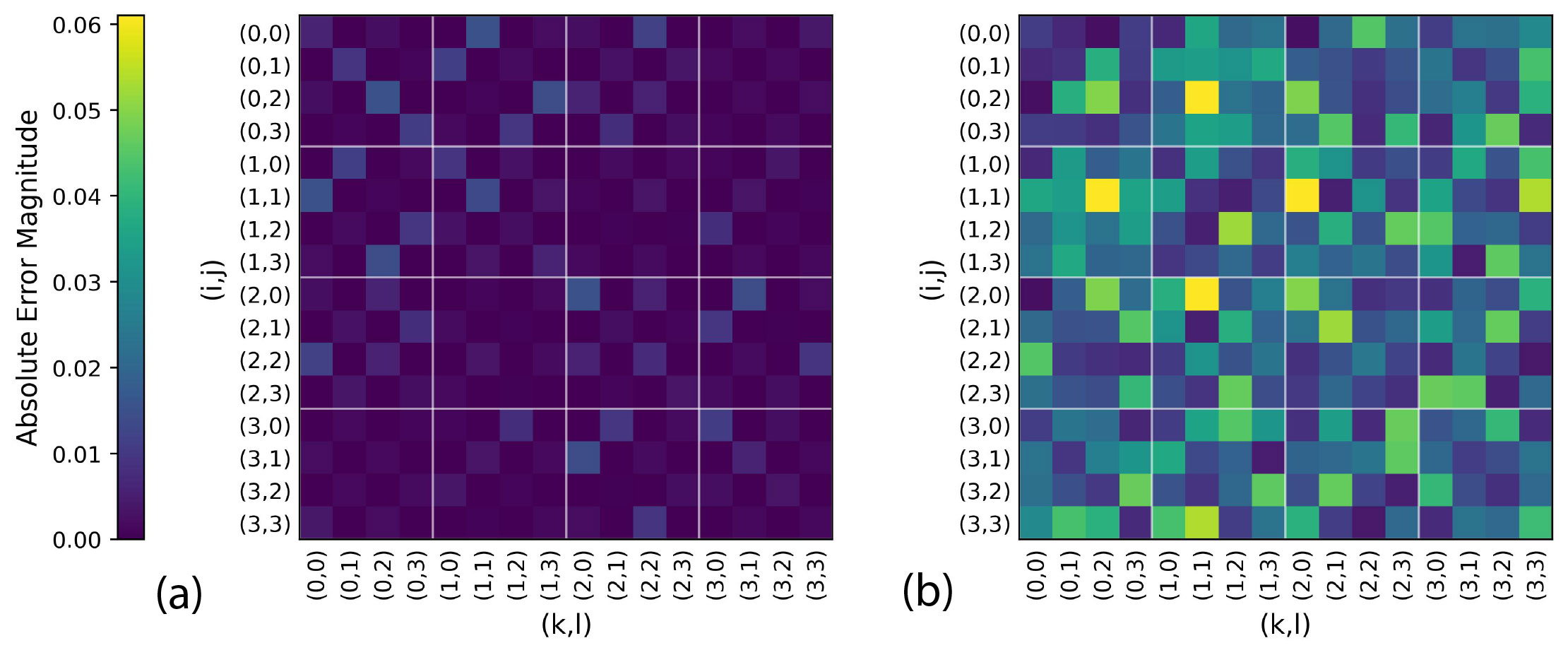}
    \caption{Comparison of absolute error matrices for the H$_4$ system in the minimal basis set obtained using the shadow variational 2-RDM (sv2RDM) (a) and Fermionic Classical Shadows (FCS) (b) methods. Each matrix element represents the absolute deviation of the reconstructed 2-RDM from the FCI reference. The color intensity denotes the error magnitude, with brighter regions corresponding to larger deviations from the reference values.}
    \label{fig:2RDM-error-matrix}
\end{figure*}

Figure~\ref{fig:N-rep} shows the lowest eigenvalue of the 2-RDM as a function of the total shot budget for H$_4$ in the minimal basis, comparing the sv2RDM method with full 2-positivity conditions (DQG) to the FCS approach, which does not inherently produce $N$-representable matrices. As the total number of shots increases, both methods converge toward consistent eigenvalue estimates, indicating improved reconstruction of the underlying fermionic state. At low shot budgets, the FCS results exhibit larger fluctuations and unphysical negative eigenvalues, reflecting the stochastic nature of randomized measurement sampling. As expected from the scaling analysis in Ref.~\cite{Low2022}, the statistical fluctuations in the FCS protocol diminish approximately as $1/\sqrt{(N_{shots})}$, leading to a smooth convergence toward the physically valid spectrum at greater sampling depths. In contrast, the sv2RDM (DQG) results remain stable across all sampling regimes with the most ``negative" eigenvalue being zero, demonstrating that explicit enforcement of $N$-representability constraints significantly mitigates noise sensitivity and maintains physically valid 2-RDM spectra.

To further address the physicality of the reconstructed RDMs, Fig.~\ref{fig:2RDM-error-matrix} compares the absolute error matrices of the 2-RDMs obtained from sv2RDM (DQG) (a) and FCS (b) for the H$_4$ system in the minimal basis. Each matrix element represents the absolute deviation from the FCI reference value. The sv2RDM (DQG) reconstruction exhibits uniformly low and smoothly distributed errors, consistent with the enforcement of $N$-representability conditions. In contrast, the FCS results display more pronounced and localized deviations, across all elements, further highlighting the statistical noise associated with such stochastic measurement sampling. The visual contrast between the two panels emphasizes that the constrained sv2RDM approach yields systematically more accurate and physically consistent 2-RDMs than the unconstrained FCS method under comparable measurement resources.

\begin{table}
\centering
\begin{tabular}{|c|c|c|c|c|}
\hline
\textbf{H chain} & \textbf{Shot Budget} & \textbf{D} & \textbf{DQG} & \textbf{FCS} \\ \hline
4 & 16,000  & 0.129 & 0.059 &  0.423 \\ \hline
6 & 36,000  & 0.226 & 0.080 & 0.873 \\ \hline
8 & 160,000  & 0.454 & 0.037 & 0.960 \\ \hline
10 & 300,000 & 0.217 & 0.055 & 1.349 \\ \hline
\end{tabular}
\caption{Comparison of Frobenius-norm differences with respect to the FCI 2-RDMs for various H chain lengths with the given total shot budgets when the semidefinite 2-RDM is optimized in the limit where $w\rightarrow\infty$ in Eq.~\ref{eq:SDP-shadow}, with partial (D) and all 2-positivity conditions (DQG) included. Shot budgets compared are: 16,000 unitaries/1 shot vs. 16 unitaries/1000 shots for H$_4$, 36,000 unitaries/1 shot vs. 36 unitaries/1000 shots for H$_6$, 160,000 unitaries/1 shot vs. 160 unitaries/1000 shots for H$_8$, and 300,000 unitaries/1 shot vs. 300 unitaries/1000 shots for H$_{10}$.}
\label{tab:hchain4}
\end{table}

Table~\ref{tab:hchain4} summarizes the Frobenius-norm differences between reconstructed and FCI reference 2-RDMs across hydrogen chains of increasing length using the same total shot budgets. The SDP used in this case has $w\rightarrow\infty$ in Eq.~\ref{eq:SDP-shadow}. The results, including only the D condition and the full DQG constraint set, are compared against the FCS protocol. Across all system sizes, imposing $N$-representability conditions systematically reduces the deviation from the reference state, with the greatest improvements observed when all DQG constraints are enforced. The effect becomes more pronounced for larger systems, where unconstrained or partially constrained optimizations show growing discrepancies from the FCS method. These results demonstrate that the inclusion of full 2-positivity conditions provides superior reconstruction fidelity, particularly under limited measurement resources and increasing system complexity.

\begin{figure}
    \centering
    \includegraphics[width=\linewidth]{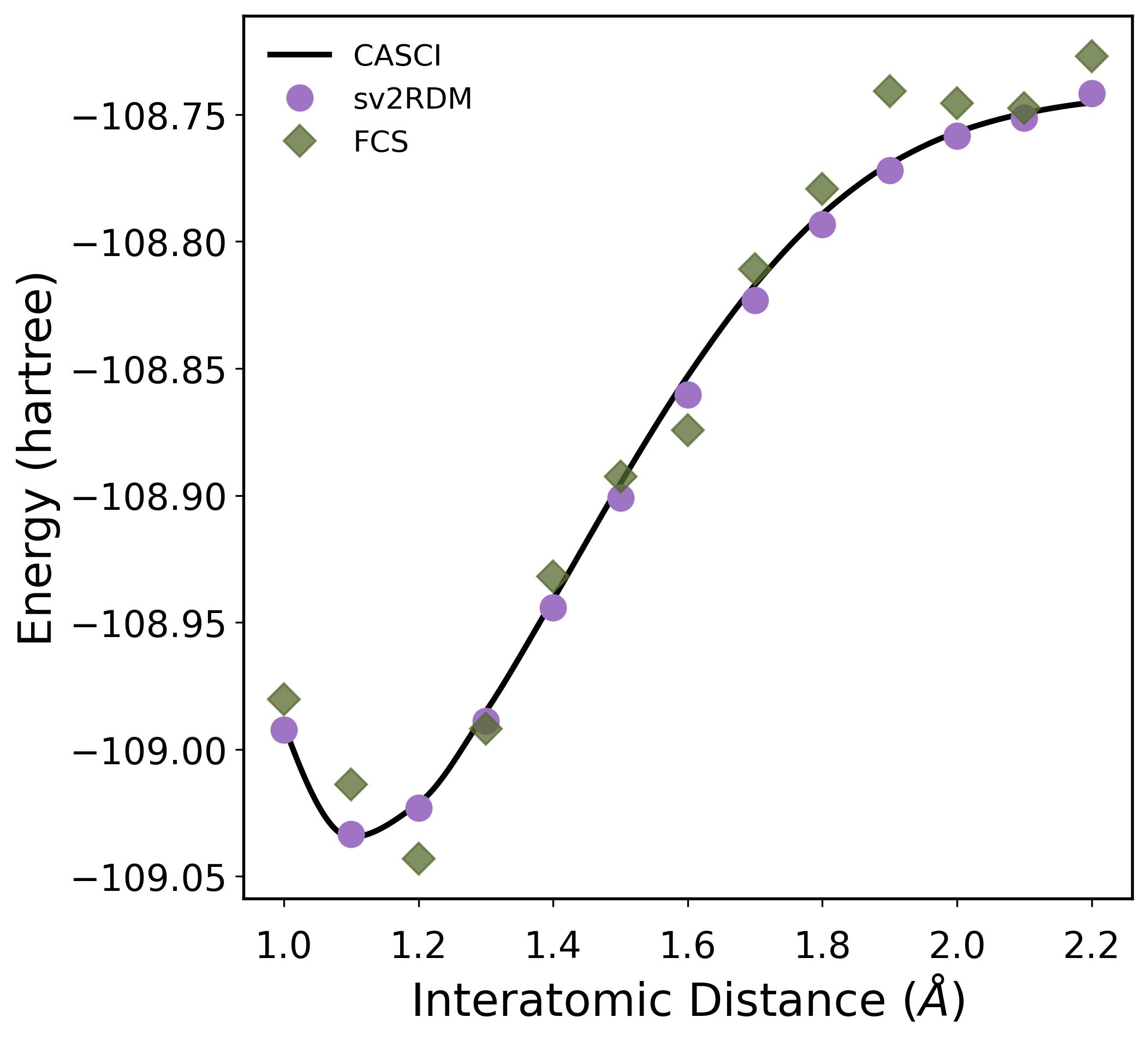}
    \caption{Potential energy curve of N$_2$ computed using the cc-pVDZ basis set and a [10,8] active space with complete active space configuration interaction (CASCI), shadow variational 2-RDM (sv2RDM), and Fermionic Classical Shadows (FCS) methods. For sv2RDM and FCS, the total shot budgets used are 100 unitaries × 1,000 shots and 100,000 unitaries × 1 shot, respectively.}
    \label{fig:N2-curve}
\end{figure}

Figure~\ref{fig:N2-curve} presents the simulated potential energy curve of N$_2$ in the cc-pVDZ basis set with a [10,8] active space using complete active space configuration interaction (CASCI), sv2RDM, and FCS methods (Supplemental Material Sec. III includes v2RDM results). The CASCI results serve as the reference for evaluating the accuracy of the reconstructed energy profiles. The FCS results exhibit noticeable deviations near the equilibrium and bond-stretching region, while the sv2RDM is able to closely reproduce reference results across the full dissociation range. These comparisons highlight that constrained semidefinite reconstructions provide quantitatively reliable potential energy surfaces even under limited measurement resources, whereas unconstrained fermionic shadow tomography approaches can struggle to maintain accuracy in strongly correlated regimes.

\begin{figure}
    \centering
    \includegraphics[width=\linewidth]{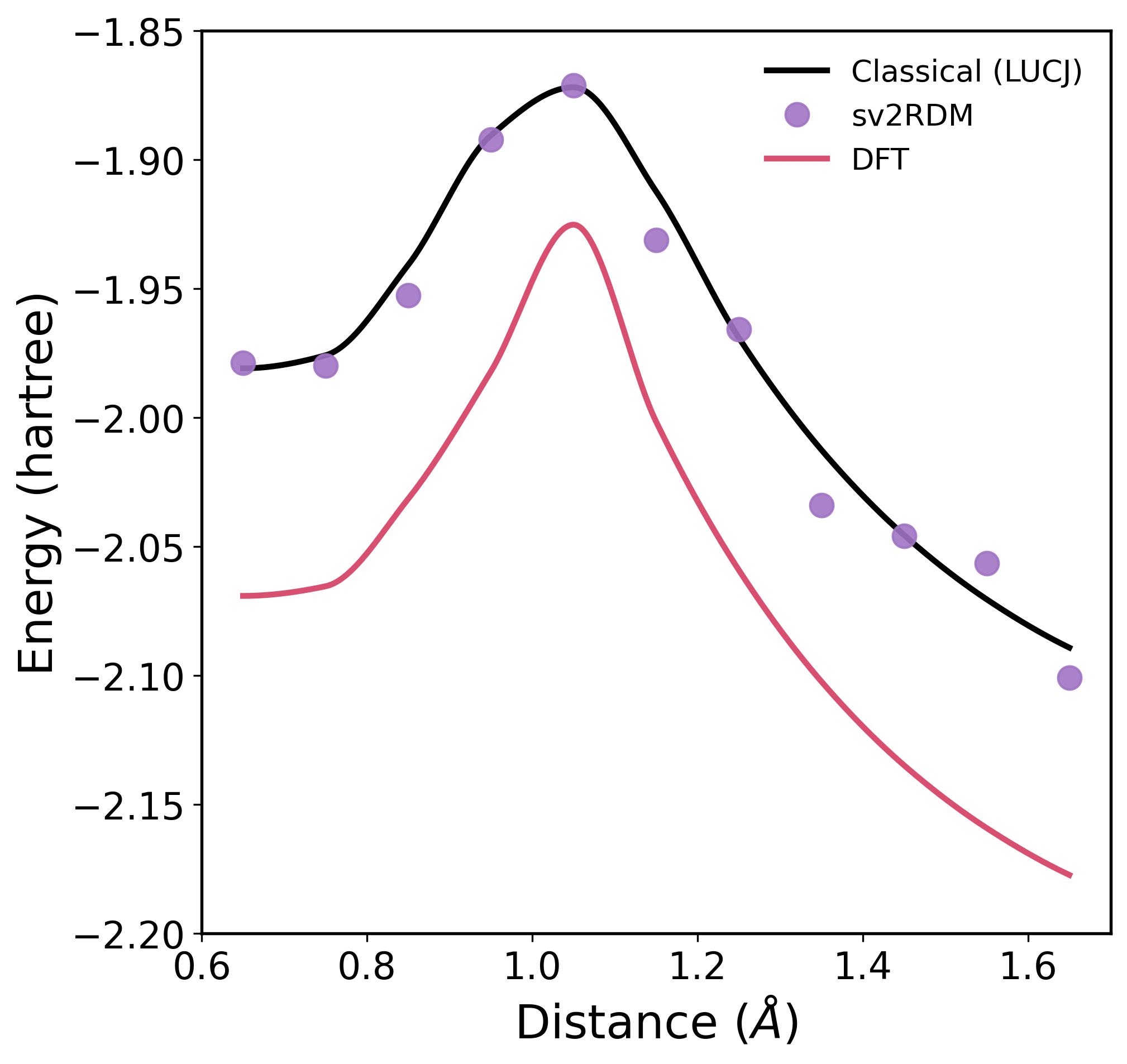}
    \caption{Potential energy curve for the H$_4$ rectangle–square–rectangle transition in the minimal basis set, computed using Density Functional Theory (DFT) with the B3LYP functional as a classical baseline, the classically computed Local Unitary Coupled Cluster with Jastrow (LUCJ) ansatz, and the shadow variational 2-RDM (sv2RDM) method computed using measurements collected on the ibm\_fez quantum computer.}
    \label{fig:H4-qcomp}
\end{figure}

To evaluate the performance of constrained shadow tomography on real quantum hardware, the H$_4$ rectangle–square–rectangle potential energy profile is computed in the minimal basis using both classical and hybrid quantum approaches (Fig.~\ref{fig:H4-qcomp}). Density Functional Theory (DFT) results using the B3LYP functional~\cite{Stephens1994, Becke1993} are shown as the classical baseline, while the sv2RDM calculations are performed using measurements collected on the ibm\_fez quantum processor (details can be found in Supplemental Material Section II A). The Local Unitary Coupled Cluster with Jastrow (LUCJ) ansatz is optimized and computed classically, serving as the initially prepared quantum state and a reference for the sv2RDM calculations. The DFT results fail to reproduce the smooth energy flattening near the square geometry that signals strong static correlation. In contrast, the sv2RDM results recover the correct qualitative features, closely matching the expected multireference character. The sv2RDM energies obtained from the quantum device align well with the LUCJ benchmark within statistical uncertainty, demonstrating that error-mitigated semidefinite constraints can effectively suppress noise-induced artifacts and maintain physical consistency in near-term quantum simulations.

\begin{table}[h!]
\centering
\begin{tabular}{|c|c|c|c|c|}
\hline
\textbf{H chain} & \textbf{sv2RDM} & \textbf{LUCJ} & \textbf{FCI} \\ \hline
4 & 0.011 & 0.034 &  0.068\\ \hline
6 & 0.013 & 0.062 & 0.101\\ \hline
8 & 0.024 & 0.061 & 0.133\\ \hline
10 & 0.065 & 0.084 & 0.166\\ \hline
\end{tabular}
\caption{Comparison of correlation energies with respect to Hartree Fock for various H chain lengths. Measurements for sv2RDM calculations are collected on the ibm\_fez quantum device with the following shot budgets: 16 unitaries/10,000 shots for H$_4$, 36 unitaries/10,000 shots for H$_6$, 160 unitaries/10,000 shots for H$_8$, and 300 unitaries/10,000 shots for H$_{10}$. Hartree Fock, LUCJ, and FCI calculations are run classically. All calculations use a minimal basis set.}
\label{tab:hchain-device-data}
\end{table}

The scalability of the constrained reconstruction approach is further evaluated by comparing correlation energies for hydrogen chains of increasing length, computed with sv2RDM, FCI, and the LUCJ ansatz in the minimal basis (Table~\ref{tab:hchain-device-data}). sv2RDM energies are obtained from measurement data collected on the ibm\_fez quantum processor under fixed total shot budgets scaled with system size, with up to 20 qubits (details can be found in Supplemental Material Section II B). Other methods were computed classically. Across all chain lengths, sv2RDM reproduces the qualitative energy trends of the reference LUCJ results, maintaining size-extensive behavior and smooth energy variation. Though deviations from the classical references appear at all system sizes, attributable to the increased noise inherent to hardware sampling, the results highlight that, even with limited quantum resources, enforcing semidefinite constraints enables systematically scalable energy predictions for chemical systems.

The results demonstrate that constrained shadow tomography provides a robust and systematically improvable framework for extracting physically meaningful quantities from both simulated and experimental quantum data. Across molecular benchmarks, enforcing $N$-representability through the 2-positivity conditions yields stable energy estimates, well-behaved eigenvalue spectra, and accurate two-particle correlations even under modest shot budgets. Comparisons with classical v2RDM, LUCJ, and FCI references confirm that the sv2RDM approach retains quantitative accuracy while remaining resilient to hardware noise and stochastic sampling errors. These findings establish a consistent foundation for assessing scalability, error mitigation, and resource efficiency in near-term quantum simulations, motivating a deeper discussion on methodological implications and potential pathways toward chemically accurate hybrid quantum-classical frameworks.

%Discussion and Conclusions
\section{Discussion and Conclusions}
The results presented in this work highlight the advantages of incorporating physical constraints into scalable quantum state reconstruction protocols. By embedding $N$-representability directly into the shadow tomography framework, our constrained shadow tomography formulation enables physically consistent and noise-resilient reconstructions of RDMs. The integration of bi-objective semidefinite programming, nuclear-norm regularization, and $N$-representability jointly suppresses measurement-induced errors, yielding accurate and stable estimates under limited shot budgets and realistic noise conditions. Benchmark results for hydrogen chains and the N$_2$ molecule confirm that enforcing physical constraints during reconstruction improves both accuracy and robustness across a range of system sizes and sampling budgets, while remaining compatible with shallow circuits. Beyond quantitative improvements, this approach provides a unified and computationally efficient framework that connects statistical inference, convex optimization, and quantum chemistry post-processing. The demonstrated reduction in measurement requirements and enhanced robustness underscore the potential of constraint-enforced tomography as a key component of hybrid quantum–classical workflows for benchmarking and validating quantum simulations of correlated fermionic systems.

The presented algorithm establishes a practical foundation for advancing quantum state characterization in many-body quantum simulations. Future work may explore the incorporation of additional system symmetries or adaptive weighting strategies to further enhance RDM reconstruction fidelity. As quantum hardware continues to progress, methods that integrate physical insight, optimization principles, and statistical learning will be essential for developing reliable and scalable approaches to quantum state reconstruction. This study positions constrained shadow tomography as a powerful and generalizable framework for extracting accurate physical insight from limited quantum data, paving the way for robust, physically grounded characterization of complex quantum systems.

%Acknowledgments
\begin{acknowledgments}
We thank Nate Earnest-Noble and Jamie Garcia for valuable discussions.  D.A.M gratefully acknowledges support from IBM under the IBM-UChicago Quantum Collaboration.  I.A. gratefully acknowledges the NSF Graduate Research Fellowship Program under Grant No. 2140001.
\end{acknowledgments}

% \appendix

\bibliography{references}

\end{document}